\documentclass[12pt,preprint]{aastex}

\newcommand{\chandra}{{\it Chandra}}

\shorttitle{Classifying X-ray Sources in External Galaxies}
\shortauthors{Prestwich et al.}

\begin{document}

\title{Classifying X-ray Sources in External Galaxies from X-ray Colors}

\author{A .H. Prestwich}
\affil{Harvard-Smithsonian Center for Astrophysics, Cambridge, MA 02138}
\author{J.A.  Irwin}
\affil{University of Michigan}
\author{R. E. Kilgard, M. I. Krauss, A. Zezas, F. Primini, P. Kaaret}
\affil{Harvard-Smithsonian Center for Astrophysics, Cambridge, MA
02138}
\author{B. Boroson}
\affil{Center for Space Research, MIT}

\begin{abstract}
The X-ray populations of Local group galaxies have been
classified in detail by Einstein, ROSAT and ASCA revealing a mix of
binaries, supernova remnants and HII regions.  However, these
observatories were unable to resolve X-ray sources in galaxies beyond
the local group.  With \chandra's exquisite spatial resolution we are
able to resolve sources in a sample of nearby galaxies.  We show that
there are highly significant differences in the X-ray colors  of
sources in bulge and disk systems.  In particular, we find that there
is a population of X-ray soft, faint sources in disk galaxies not seen
in bulges, and a smaller population of hard sources also seen
preferentially in disk systems.  These differences can  be used as a basis to classify sources as low and high
mass X-ray binaries, supernova remnants and supersoft sources. We
suggest that the soft sources seen preferentially in disks are
probably dominated by supernova remnants, although we cannot rule out
the possibility that they are a new population of absorbed, faint,
supersoft accretion sources associated with the young stellar
population.  The hard sources seen in disks but not bulges we identify
as high mass X-ray binaries.   While
it is impossible to classify any individual source on
the basis of X-ray color alone, the observations presented here
suggest that  it is possible to separate sources
into groups dominated by one or two source types.  This 
classification scheme is likely to be very useful in population studies, where it is crucial to distinguish between
different classes of objects.

\end{abstract}

%\keywords{surveys --- galaxies: spiral ---
%galaxies: starburst --- X-rays: galaxies --- X-rays: sources}

\section{Introduction}

\label{intro}
Our galaxy  and other nearby galaxies contain a multitude of X-ray
sources \citep{fab89}.  Observations of sources in the Milky Way  and  Local Group galaxies
show that many of the brightest X-ray 
sources are  high and low mass  X-ray binaries (HMXB, LMXB), but there are
also sources associated with supernova remnants (SNR) and HII regions.
For example, the ROSAT survey of M31 \citep{supper01} revealed a total of 560 X-ray
sources, 16 of which have been optically identified with supernova remnants and 33
with globular clusters.  Most of the bulge sources in M31 have been
identified on the basis of their X-ray spectra as low mass X-ray binaries  \citep{trinch99,primini93}.  Studies
of X-ray populations in a wide range of galaxies is potentially very
valuable.  For example, since X-ray binaries are products of the end point of
stellar evolution, the X-ray binary population  should be related to
the history of star formation in each galaxy \citep{wu01,rek02_1}.  However, the resolution of {\it Einstein}, ROSAT and ASCA
was not sufficient to allow
identification of sources in galaxies beyond the Local Group. \chandra\ has the angular resolution necessary to study the X-ray source
population in more distant galaxies for the first time
\citep{blanton01,tenn01,irwin02,fab01,soria02}.

In this paper we compare X-ray populations of a sample of disk and
bulge galaxies.  We show that there is a highly significant difference
in the X-ray colors of sources in bulge and disk systems.  Disk
galaxies have a population of soft sources and a population of hard
sources not seen in bulges.  X-ray colors
are known to be a sensitive discriminator of source type
\citep{white84,haberl00,yok00,sasakil00} and we interpret these differences as
differences in the source population.  Furthermore, we suggest that X-ray colors derived from \chandra\ observations can be used to identify
different classes of X-ray sources in nearby galaxies.  In Section~\ref{anal} we
describe our data analysis. In Section~\ref{sec:b_and_d} we show
the difference in the distribution of
X-ray colors in bulge and disk systems and in Section~\ref{sec:id}  we
suggest a classification scheme based on colors. In
Section~\ref{sec:abs} we discuss the effects of absorption on source
classification, and in Section~\ref{sec:LVSD} we
discuss the spatial distribution, luminosities and variability
characteristics of the different classes of sources.  Finally in Section~\ref{sec:s_and_c} we summarize our results
and give suggestions for further work.

\section{Data Analysis}
\label{anal}

Our sample for this study consists of 2 late-type face-on spiral galaxies
(M101 and M83) 
and 3 elliptical/bulge systems (NGC 4697, the inner bulge of M31 and
the bulge of NGC 1291).  
Observations of all galaxies were performed on
the back-illuminated ACIS-S3 CCD. All analysis was performed using the \chandra\ Interactive
Analysis of Observations software package ({\it CIAO}) v2.2.1
and the \chandra\ Calibration Database v2.10.  Data were
screened for times of high background, as many of the
observations occurred near periods of solar activity. Exposure
times of the screened observations are in
Table~\ref{tab:prop}.  Source lists were constructed using
{\it wavdetect} \citep{pef02}, the Mexican-hat wavelet source
detection routine which is part of {\it CIAO}.  We ran {\it
wavdetect} on 0.3-6.5 keV band images using wavelet scales of
2, 4, 8, and 16 pixels.  This combination of energy band and
wavelet scales yielded the fewest spurious detections and
found all visually obvious point sources. The default value of the
parameter {\tt sigthresh} ($1\times10^{-6}$) was used to identify
source pixels. The output source
regions were visually inspected to identify multiple detections and single sources detected twice.    Nuclear sources were
removed from the source lists.  Sources  outside of the ACIS-S3 chip
were excluded from further analysis.  Sources outside the optical
boundaries of the galaxy, defined as the observed diameter at 25 mag
arcsec$^{-2}$ in the blue band, $D_{25}$,  were also excluded, since these have a high probability of being background
objects.  Values for $D_{25}$ were taken from \cite{tully88}.
Table~\ref{tab:prop} gives the total number of sources detected in
each galaxy with minimum of 20 net counts  and  significance
$>$4$\sigma$ (see Section~\ref{sec:b_and_d}) and the expected number of
background sources (same significance) from the logN-logS curves of
\cite{giacconi01}.  Variations in the number of background sources with position on the
sky \citep{cowie02} will change the number of expected background 
sources by at most a factor of 3.

The total number of counts for each detected source were calculated in
3 bands: the soft band (0.3-1 keV), medium band (1-2keV) and hard band
(2-8keV).   Counts were extracted from source regions determined by 4-$\sigma$ source ellipses
from {\it wavdetect}.  This over-estimates the source size, but the
contribution from background is negligible.
The background region for each source was taken to be an ellipse with major and minor axes equal to 4
times the source axes and excluding the source region and
any other overlapping source regions.  Background ellipse radii were
not allowed to exceed 50 pixels (about 25\arcsec) to avoid
contamination from variations in the diffuse emission.
Source count rates were corrected using monochromatic exposure maps
created for each band using a monochromatic response at the mean
photon energy in each band:
0.65 keV (soft band), 1.5 keV (medium band) and 5 keV (hard band).
These maps take into account vignetting and spatial variation of the
CCD QE.   X-ray colors were then calculated for each source as:
$H1=(M-S)/T$ (soft),
 and $H2=(H-M)/T$ (hard),  where $S$, $M$ and $H$ are the
counts in the soft, medium and hard bands respectively, and $T$ is the
total counts in all three bands combined.

We estimated fluxes from each source  assuming a 5~keV thermal
bremsstrahlung model with 
photoelectric absorption.
The $n_H$ was fixed at the Galactic value. No correction for
absorption in the host galaxy was applied.  
Luminosities were derived in the 0.3-8 keV band using distances from the Nearby
Galaxies Catalog (Tully, 1988), which assumes $H_o=75$ km
s$^{-1}$ Mpc$^{-1}$ coupled with a Virgo infall model.  A more
detailed description of the data analysis, discussing the relevant
biases and presenting source lists with spectral and timing results
will be presented in \citet{rek02}.

\section{Color differences between Bulge and Disk Sources}
\label{sec:b_and_d}

\begin{figure}
 %\epsscale{0.5}
 \plotone{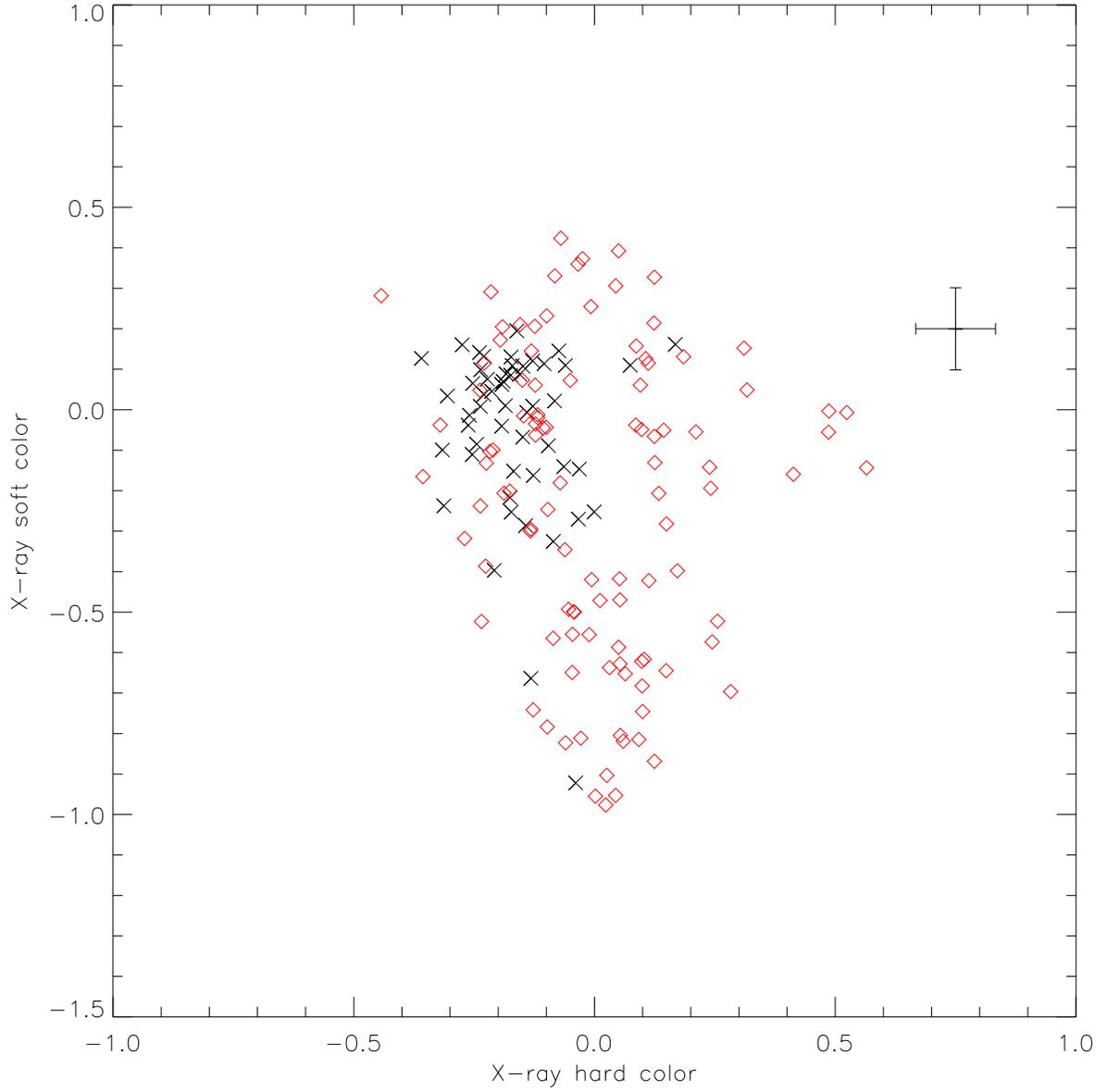}
\caption{X-ray hard color (X-axis) plotted again X-ray soft color for
the inner bulge of M31 (X's) and the disk of M101 (diamonds). . X-ray
hard color is defined as  $H2=(H-M)/T$, X-ray soft color as  $H1=(M-S)/T$ }
\label{fig:m31_and_m101}
\end{figure}

Figure~\ref{fig:m31_and_m101} shows the X-ray color-color diagram for
sources in the
inner bulge of M31 and the disk of M101.    Only
sources with a minimum of 20 net counts  and  significance
$>$4$\sigma$ are plotted and used in subsequent analysis. 
There is a clear difference in the colors of sources in M31 and M101.
The bulge sources in M31 are clustered in a relatively narrow
color range, with both H1 and H2 having values between -0.4
and 0.4.  The M101 disk sources, on the other hand, have a tail of soft
sources extending to $H1=-1$.  M101 also has a
population of sources  with $H1>0.1$, not seen in M31.  It is
important to note that the  M31 exposure is deeper than the M101
observation (see Table~\ref{tab:prop}).  The approximate luminosity limit for sources
with $\ge 20$ counts in M31 is nearly a factor of two lower than for
sources in M101.   Hence if the ``extra'' soft and hard sources
existed in M31 they should have been detected.

\begin{figure}
 %\epsscale{0.5}
 \plotone{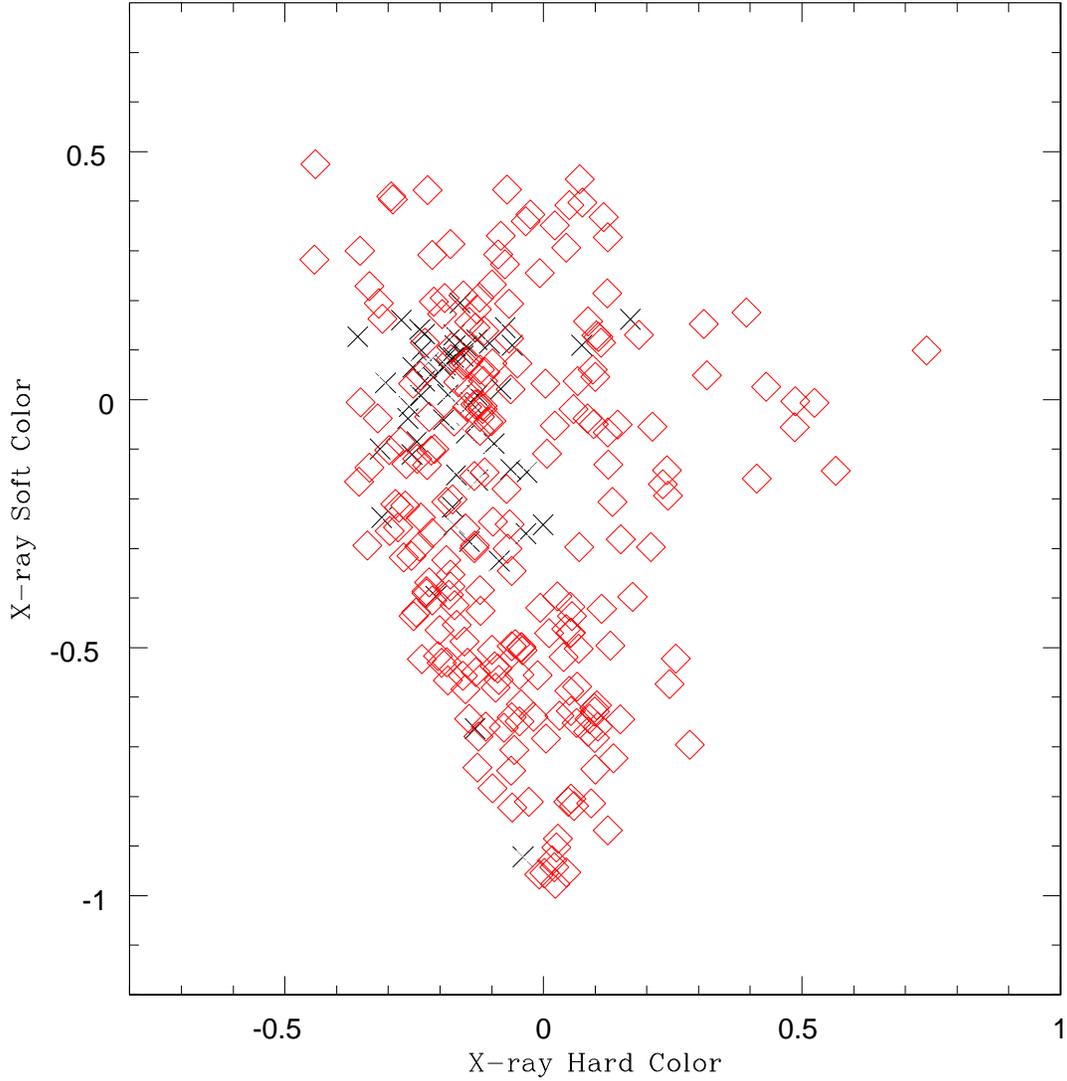}
\caption{X-ray hard color (X-axis) plotted again X-ray soft color for
 a selection of bulge and disk systems.  The red diamonds are sources
from the disk galaxies M101 and M83 and the black X's are sources
from bulge systems (inner bulge of M31, bulge of NGC 191 and the
elliptical NGC 4697) . X-ray
hard color is defined as  $H2=(H-M)/T$, X-ray soft color as  $H1=(M-S)/T$ }
\label{fig:b_and_d}
\end{figure}

Figure~\ref{fig:b_and_d} shows the X-ray color-color diagram for
sources drawn from  a
larger sample of galaxies (the bulges of M31 and NGC 1291, the
elliptical galaxy NGC 4697 and the disk galaxies M83 and M101.)  The
sources were selected according to the criteria described in the
previous paragraph.
The color differences between disk
and bulge sources are in this larger sample is very clear, and highly significant.  A Kolmogorov-Smirnov (K-S) test
gives the probability that the two data sets arise from the same
distribution as $2.5\times10^{-7}$ (soft color) and $9.5\times10^{-5}$
(hard color).

\section{Classification of Sources}
\label{sec:id}

The differences in the colors between bulge and disk sources indicates
a difference in the source population, and can be used classify the
sources in different systems.  While it is impossible to classify any
individual source  with confidence on the basis of X-ray colors alone,
it is valuable to be able to separate sources in a statistical sense
for population studies.  

\begin{figure}31
 %\epsscale{0.5}
 \plotone{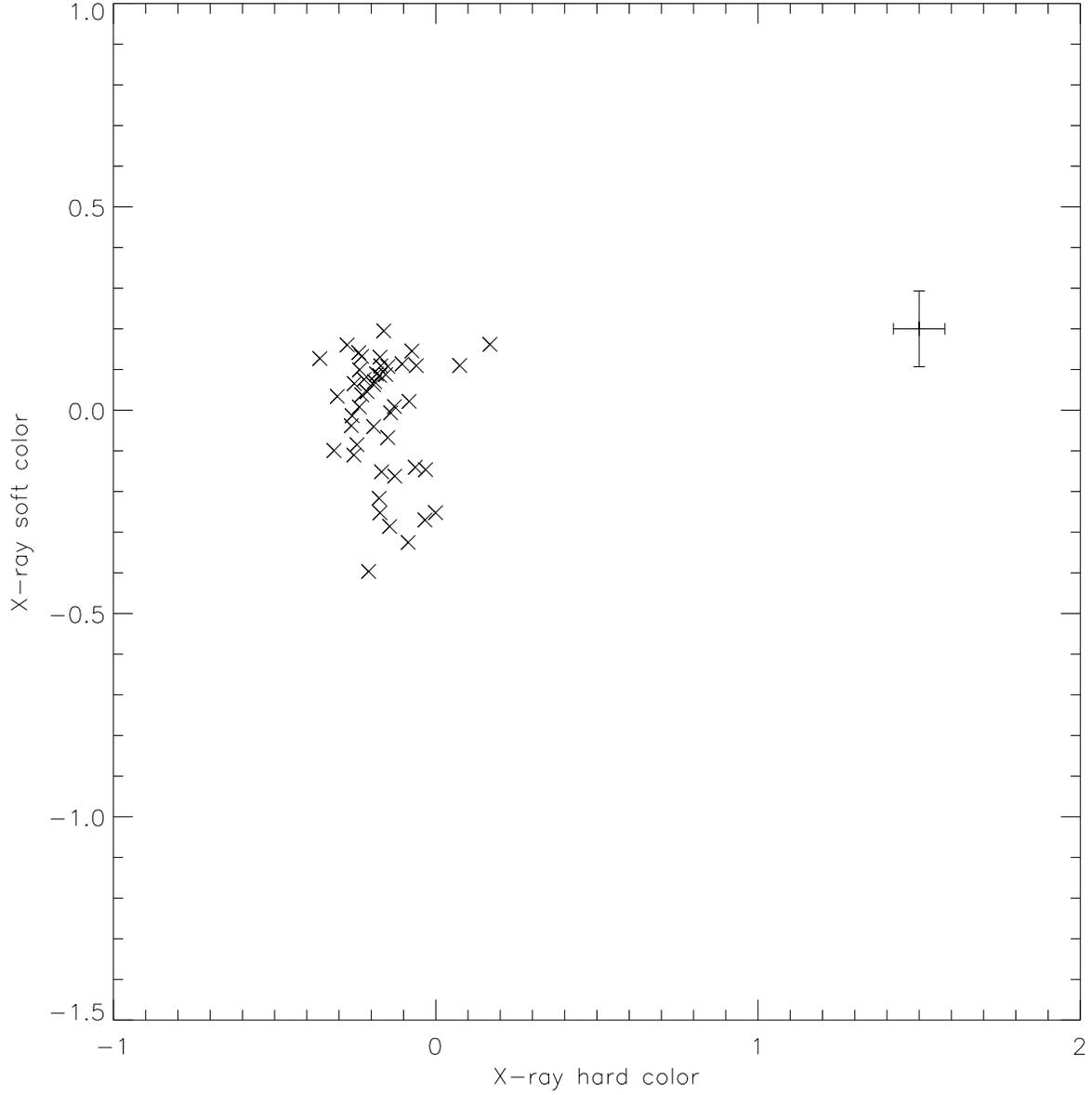}
\caption{X-ray color-color diagram for LMXBs in M31.   X-ray
hard color is defined as  $H2=(H-M)/T$, X-ray soft color as  $H1=(M-S)/T$ }
\label{fig:xcol_lmxbonly}
\end{figure}

\begin{figure}
 %\epsscale{0.5}
 \plotone{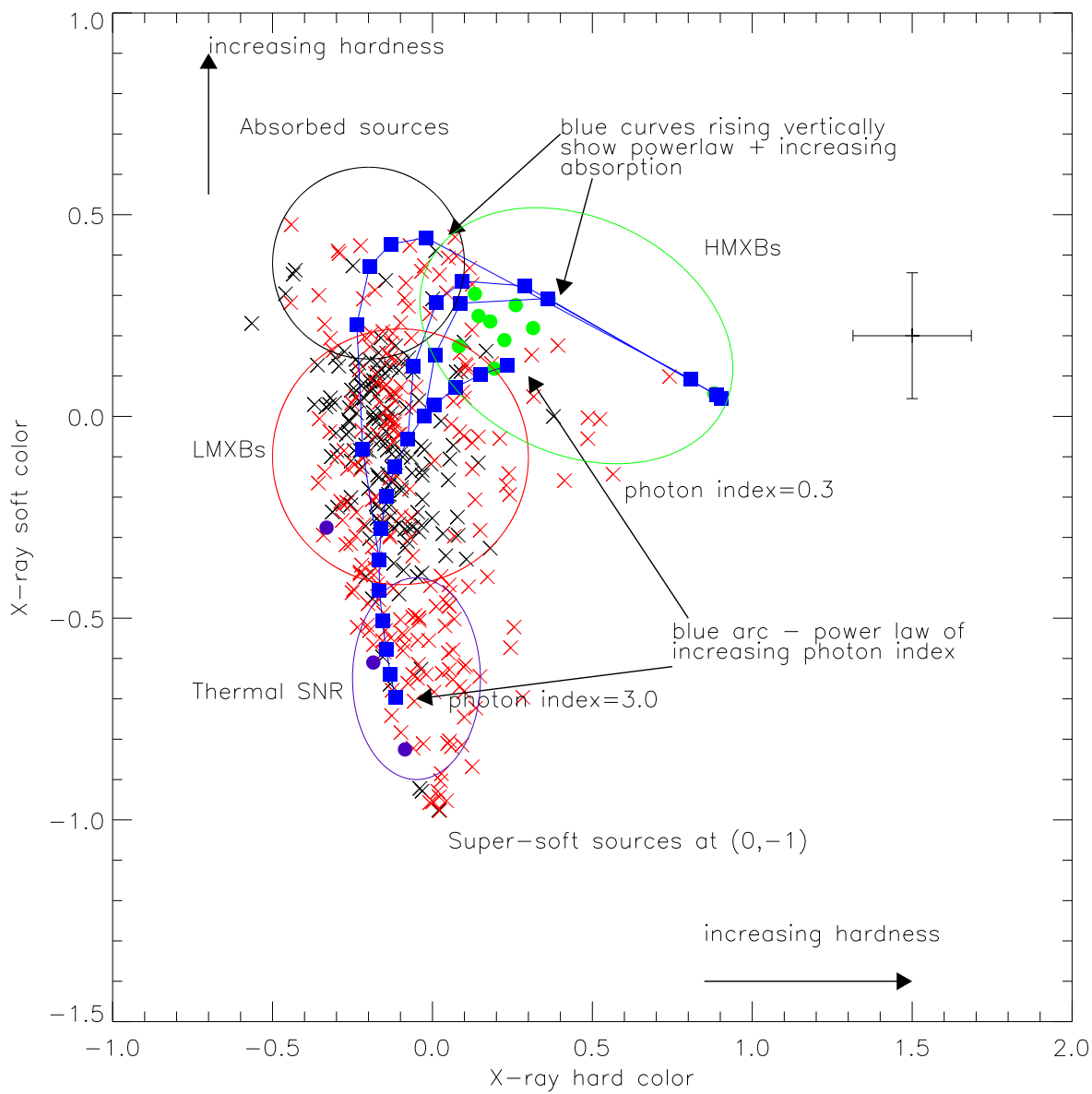}
\caption{Proposed classification scheme.
The red circle outlines the area of the color-color diagram which is
probably dominated by LMXBs, the yellow ellipse the region occupied by
thermal SNR, and the green ellipse the region occupied by HMXBs.  The
black circle delineates the most absorbed sources, and is probably a
mixture of types.  The blue arc of points stretching from (0.3,0.15)
to (-0.1,-0.7) show the colors of a simple power law spectral model
with photon index increasing from 0.7 to 3.0.  The blue curves rising
vertically show the effect of adding absorption to power law slopes of
photon index 1.0, 1.2 and 2.0.}
\label{fig:xcol_models}
\end{figure}

As
mentioned in Section~\ref{intro}, the bulge of our Galaxy and M31 as
well as sources in elliptical galaxies are dominated by low mass X-ray
binaries \citep{fab89,grimm01}.  We therefore suggest that the region of the X-ray
color-color diagram populated by ``bulge'' sources  contains many
LMXBs.   The
``bulge'' sources are in a region of the diagram characterized by a
simple power law slope (photon index 1-2.5)  with some intrinsic (to the source)
absorption.  This is typical of LMXBs \citep{white95}.  The ``LMXB'' region is best defined in
Figure~\ref{fig:xcol_lmxbonly}. By cross-correlating  our M31
inner bulge source list with the list given by \cite{kong02} we have 
rejected known foreground and background sources, supersoft sources, and sources for
which there may be a contribution from hot gas (e.g. those associated
with planetary nebula).  The resulting  plot is
illustrative of the intrinsic dispersion in the color-color diagram of 
LMXBs.  Very few of the soft or hard sources described in
Section~\ref{sec:b_and_d} and associated with disk regions are likely
to be LMXBs.  The LMXB area is shown in Figure~\ref{fig:xcol_models}.
Figure~\ref{fig:xcol_models} also shows the predicted colors of
power-law spectra with increasing photon index (blue arc).
 The effect of adding absorption to a
simple power law is illustrated with the vertically rising blue lines.

The soft sources ($H1$ between -0.4 and -0.9)  which appear
almost exclusively in disks may be  identified as thermal
supernova remnants.  Supernova remnants are found in regions of
on-going star formation, and hence very few are found in elliptical galaxies.  Thermal supernova
remnants also have soft spectra \citep{long96} consistent with the colors seen
here.  This is demonstrated in Figure~\ref{fig:xcol_models}.  The
purple boxes show the colors of several known thermal supernova remnants in
the Small Magellenic Cloud with spectra measured by ASCA \citep{yok00}.  Most of
these sources lie below the ``LMXB'' region, in the part of the
diagram populated by by soft sources.  One absorbed SNR has $H1=-0.3$.  Additional evidence that these sources are
indeed thermal SNR comes from a detailed study of the discrete source
population in M83 by \cite{soria03}.  They find that two of the
brightest soft sources have X-ray spectra dominated by emission lines,
typical of SNR.  However, it is important to stress that
some of these soft disk sources may be absorbed supersoft sources
(see discussion below.) SNR dominated by non-thermal emission
(crab-like objects) may also contribute to the source population.
These will have spectra somewhet harder than thermal SNR, and will
probably be located in the ``LMXB'' or ``absorbed sources'' part of the
diagram.

There are a handful of sources with $H2>$0, which  appear in the disks but not in the bulges.  The most natural
interpretation of these sources is that they are high mass X-ray
binaries.  High mass X-ray binaries have high mass (hence short lived)
secondaries and are typically found in regions of active star
formation. They are known to have hard spectra in the 1-10 keV region,
with a power-law photon index of 1-2, and often high
variable intrinsic absorption \citep{white95}.  The green circles in
Figure~\ref{fig:xcol_models} show the colors of known binary
pulsars observed by ASCA \citep{yok00}.   As noted above, most known HMXBs are
associated with pulsars in binary systems.  Therefore, the HMXBs shown in
Figure~\ref{fig:xcol_models} represent only pulsed (neutron star) binaries and not black
hole binaries with early-type companions.  These sources have spectra
similar to black hole LMXB companions \citep{vanP99}, and probably are
not well separated from black hole LMXB sources in the color-color diagram.

There are a small number of sources with H2=0 and H1=-1.  These sources essentially have no counts above 1keV, and
some are undoubtedly classical supersoft sources \citep{greiner91}.  These
``supersoft'' sources seem to occur in both bulge and disk systems.
They are frequently brighter than the sources identified as supernova
remnants, and most are variable (see Section~\ref{sec:LVSD}).
\citet{pence01} have identified 10 sources in M101 as supersoft.  We
do not include 6 of these fainter sources because they have less than
20 counts.  We detect the remaining four sources, three of which have
 $H1\le -0.9$ and one of which has $H1=-0.81$.  It is also possible
that some (or even all) of the soft sources identified as thermal SNR
are absorbed supersoft sources.  Absorption will decrease the flux in
the lowest energy \chandra\ band, shifting the soft color vertically
upwards on the color-color plot (see also~\ref{sec:abs}).  If this is
the case, we have identified a new population of faint, absorbed,
accretion-powered soft sources in disk galaxies, probably associated
with the young stellar population.  We favor the hypothesis
that the soft source population is dominated by SNR, but a new
population of accretion-powered soft sources cannot be ruled out.  An
in-depth study of the variability and spectral characteristics of the soft sources
(see Section~\ref{sec:LVSD}) is required to distinguish between these
two models.  

\section{Absorption and Overlap in Source Type}
\label{sec:abs}

There will undoubtedly be some overlap in source type in the
color-color diagram, both because different sources have intrinsically
similar spectra (e.g. black hole binaries with high and low mass
companions) and because of absorption. 

Photons in the lowest energy \chandra\ band will be preferentially
removed in an absorbed source, making $H1$ ``harder''.  This will move
the source vertically up the color-color diagram, as illustrated in
Figure~\ref{fig:xcol_models}.  When the absorption becomes very
severe ($n_H\sim 10^{22}$) photons in the medium \chandra\ band will
also be removed, causing $H2$ to become harder.  The source will then
curve to the right in   Figure~\ref{fig:xcol_models}, as shown by the
tracks of increasing absorption.  Absorption will cause supersoft
sources to be confused with SNR, move SNR into the LMXB region of the
diagram, and blur the distinction between LMXBs and HMXBs.  Color
information in the \chandra\ bands will be very limited for highly
inclined galaxies, where essentially all low energy photons are lost.
In this case, sources distinguished primarily on the basis of their
soft color will be confused (SNR, supersoft sources, LMXBs), and only
a very rough separation on the basis of hard color will be possible.

\section{Luminosities, Variability and Spatial Distribution}
\label{sec:LVSD}

\begin{figure}
%\epsscale{1.5}
\plottwo{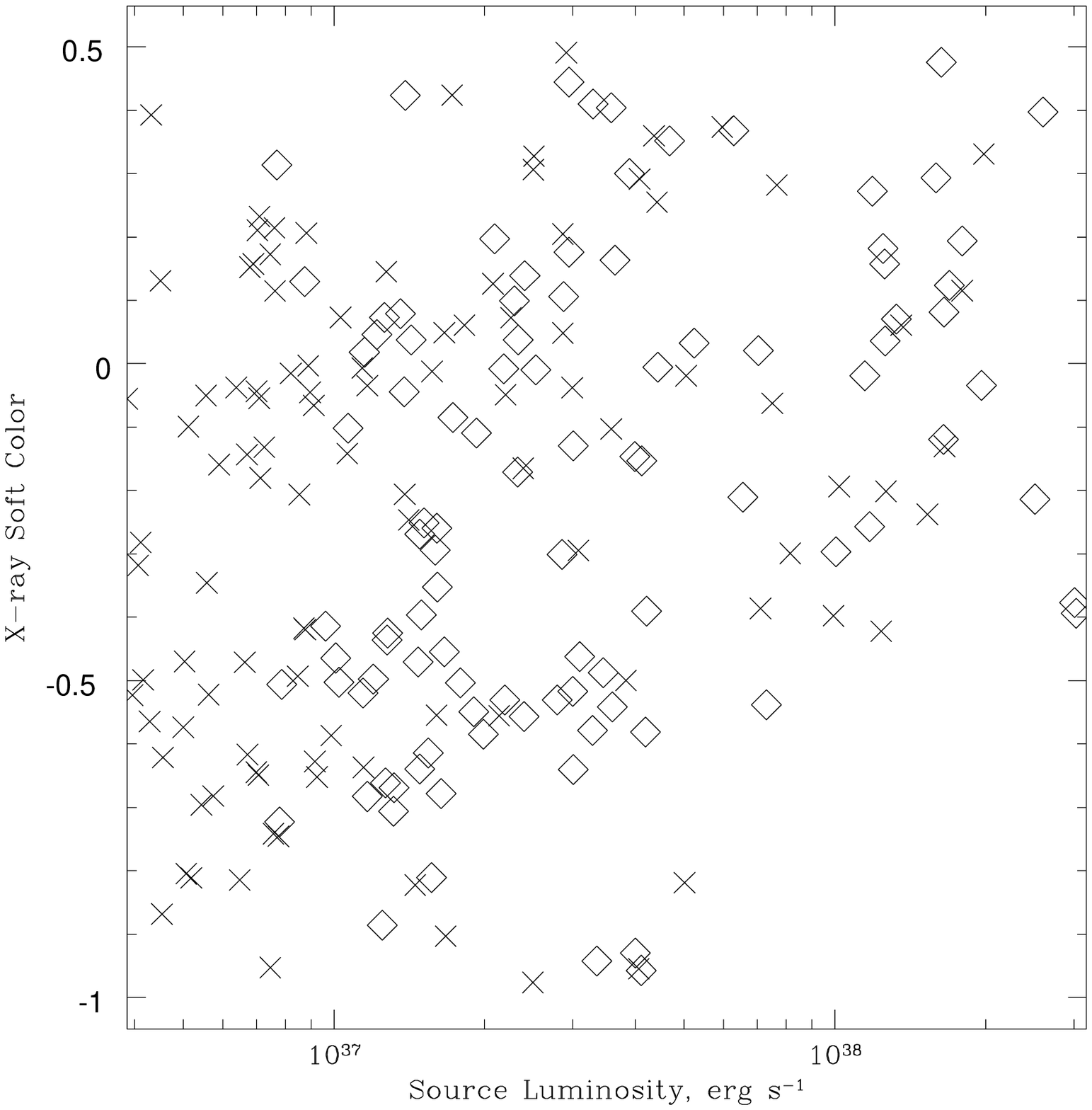}{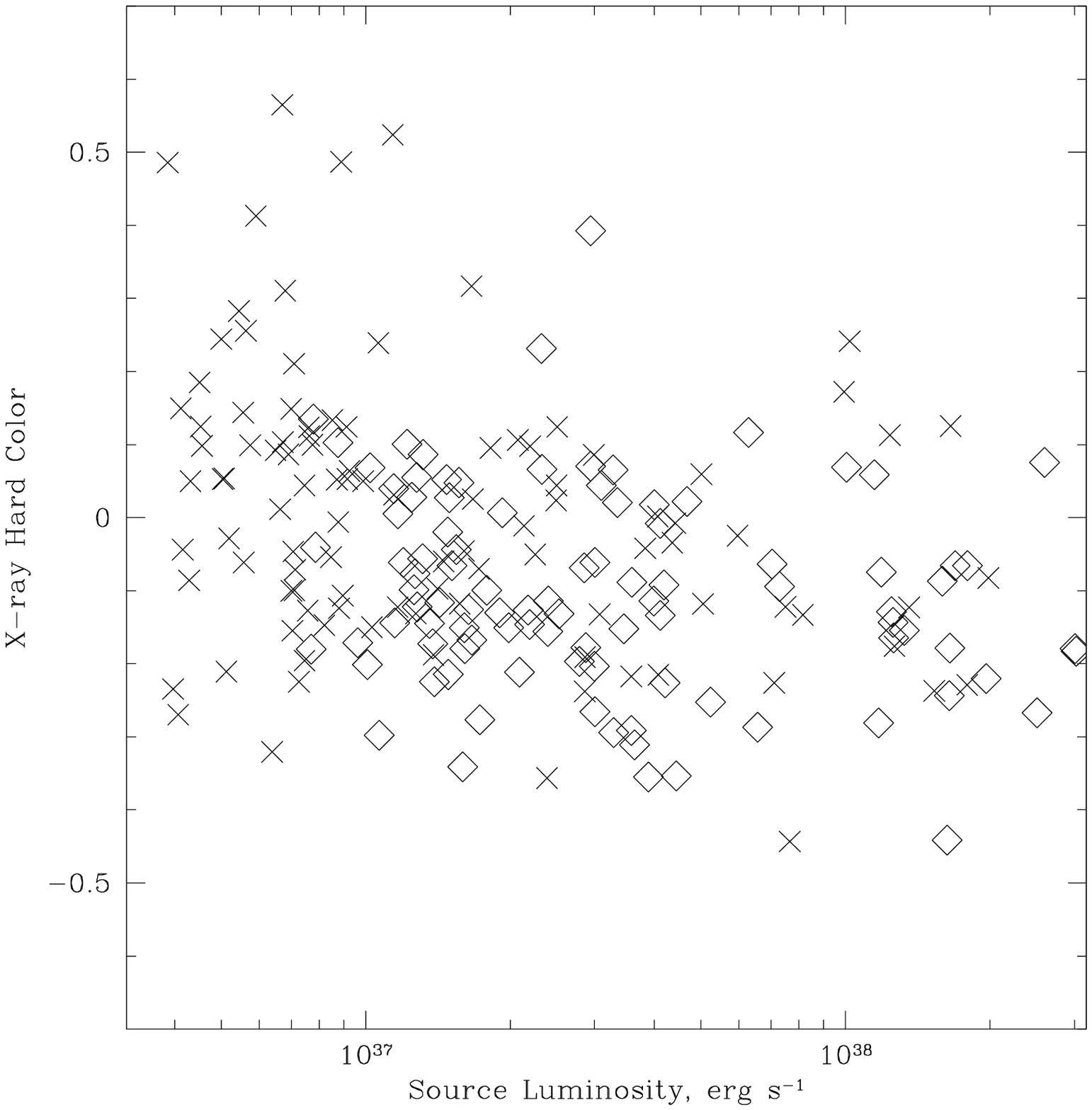}
\caption{Source luminosity plotted against X-ray soft color (left) and
X-ray hard color (right) for M101 (X's) and M83 (diamonds)}
\label{fig:colors_lumin}
 \end{figure}

The soft sources identified in Section~\ref{sec:id} as supernova
remnants have considerably less scatter in their luminosities than do
sources with $H1>$-0.6 -- i.e. a higher fraction of the sources in the
LMXB part of the diagram have luminosities  $> 10^{37}$ erg s$^{-1}$.   This is demonstrated in
Figure~\ref{fig:colors_lumin}.  The left panel shows the soft X-ray
color  plotted as a function of luminosity for M101 and M83.  This
difference in the distribution of luminosities in LMXB and ``soft''
sources is significant; a KS test gives the probability that they are
drawn from the same distribution as $6\times10^{-4}$.
The larger scatter in the luminosities of sources with $H1>-0.5$ is
naturally explained if many of these objects are accreting binaries.
Binaries can reach much higher luminosities (especially in a flare
state) than is typically observed in evolved SNR ( $10^{36}$-$10^{37}$
erg s$^{-1}$).  The X-ray luminosities of the soft sources are typical of
brighter SNR; there are certainly other soft sources below our
detection threshold.  Several of the brightest soft sources have very extreme
colors (H1=-1) and have colors and luminosities characteristic
of supersoft sources \citep{kah94}.  These are probably accretion powered.

The right panel of Figure~\ref{fig:colors_lumin} shows the luminosity plot for X-ray
hard color.  This plot suggests that sources with the hardest colors
($H2\ge$ 0.2, HMXB candidates)
have smaller scatter 
than less extreme sources.  A K-S test shows that the luminosity
distributions of the hardest sources is significantly different from
those in the LMXB part of the diagram (the probability that they are
drawn from the same distribution is $5\times10^{-5}$.  The luminosity function of LMXB
sources in the Milky Way extends to higher luminosities than the HMXB
luminosity function \citep{grimm01}, consistent with what is
observed here.  We note, however, that extremely luminous sources
tentatively associated with HMXBs
are seen in starburst galaxies \citep{zez02,fab01,prestwich01}.

If the soft sources discussed in Section~\ref{sec:id} are supernova
remnants, they should show little or no evidence for variability. 
When SNR are young ($\le$ 1000 years) several emission mechanisms might
contribute to the X-ray emission, and the flux may be variable
\citep{schl95}.  However, once the remnant enters the adiabatic phase
the luminosity should decline gradually \citep{JSS81, ham83}.   In
contrast, accretion-powered supersoft sources are known to be variable 
(\citep{kah94,kong02}, and detection of variability in a large fraction of the soft
sources would support the hypothesis that they are accretion-powered.   Both M101
and M83 have one deep exposure in the \chandra\ archive, with a second
shorter observation (see Table~\ref{tab:prop}).  The second shorter
exposure of M83 was taken 16 months after the first, and the second
exposure of M101 taken 7 months after the first.  In the long (100ks)
pointing of  M101 there are 28 soft sources (potential SNRs) detected.  Most of these have
20-100 counts and are not detected in the short (10ks) observation.
This does not provide very stringent constraints on the variability of
the soft sources, since to be detected in the short observation they
would have to increase in flux by factors of 5-10.  There
are two ``soft'' sources detected in both the long  and short
observation.  One source with $H1=-0.8$ shows no evidence
for variability.   Another source is significantly variable, and has a
$H1=-0.78$ in the long observation.  This source has a
luminosity of $\sim10^{39}$ erg s$^{-1}$, and is the brightest object in
M101 (source 98 from \citet{pence01}).  It is highly variable and is
clearly an accretion source.  Two ``hard'' (possible HMXB) sources are detected in
both observations in M101.  One has $H2=0.31$ in the long observation and shows no evidence for variability.  One source has $H2=0.48$, and the flux approximately doubles in the second
observation.

There are 28 soft (possible SNR)  sources detected in M83 in the long (50ks)
observation.    A total of 7 sources with soft colors that were
detected in the long observation were also detected in the short observation.  Two of these are clearly variable,
while the remaining 5 have approximately constant flux. Of the 28 soft
sources detected in the long observation,  5
sources should have been detected in the short (10ks) observation,
assuming no variability.  Three of these have  extremely soft colors,
are clearly variable and are probably supersoft sources.  The
remaining two sources may have declined in flux between the two
observations.   Three hard
(potentially HMXB sources) were detected in the long observation of
M83.  One of these sources may be variable.  We conclude that we have
insufficient data to make definitive statements about the variability
properties of the soft sources in either M83 or M101.

\begin{figure}
\epsscale{1.5}
\plottwo{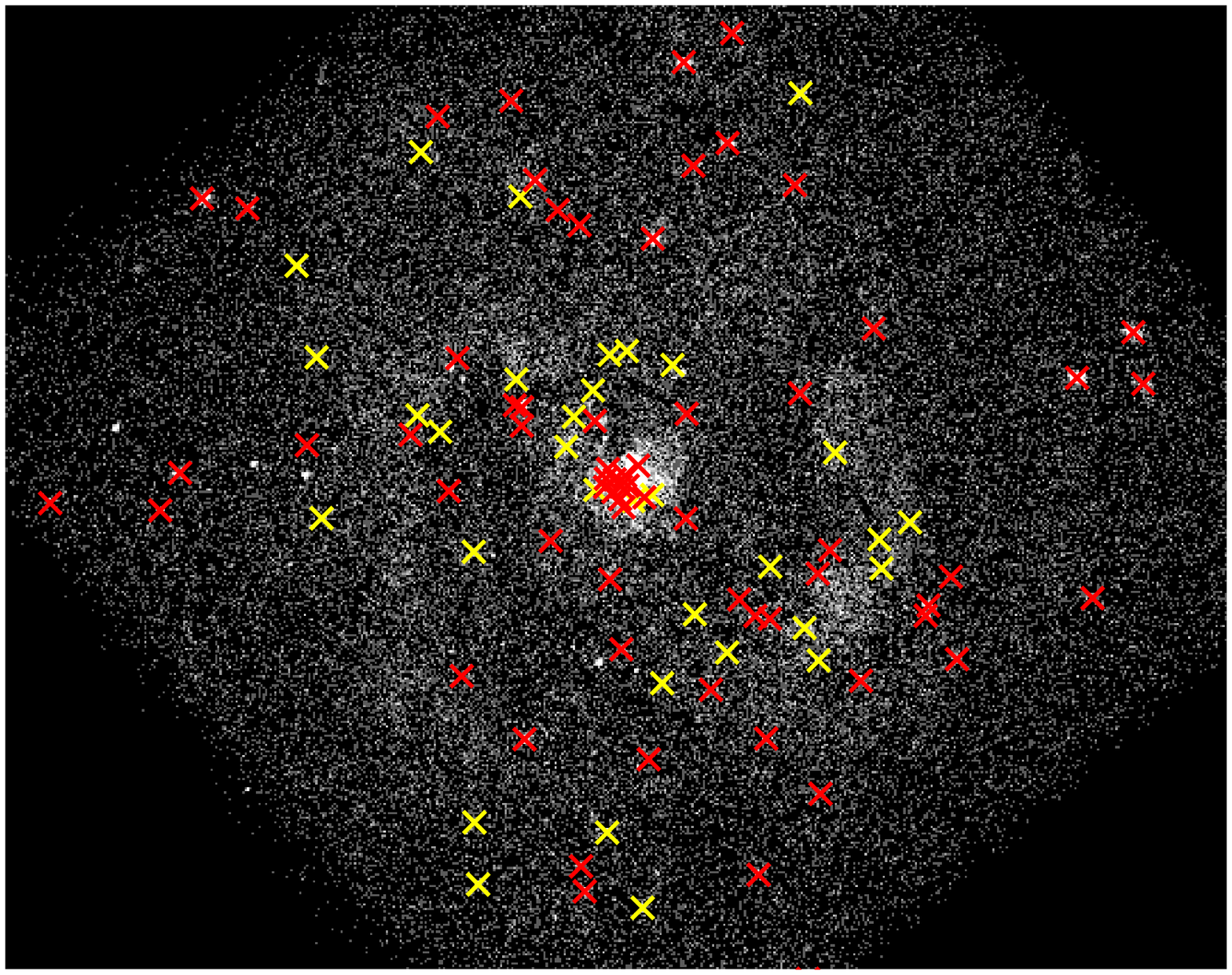}{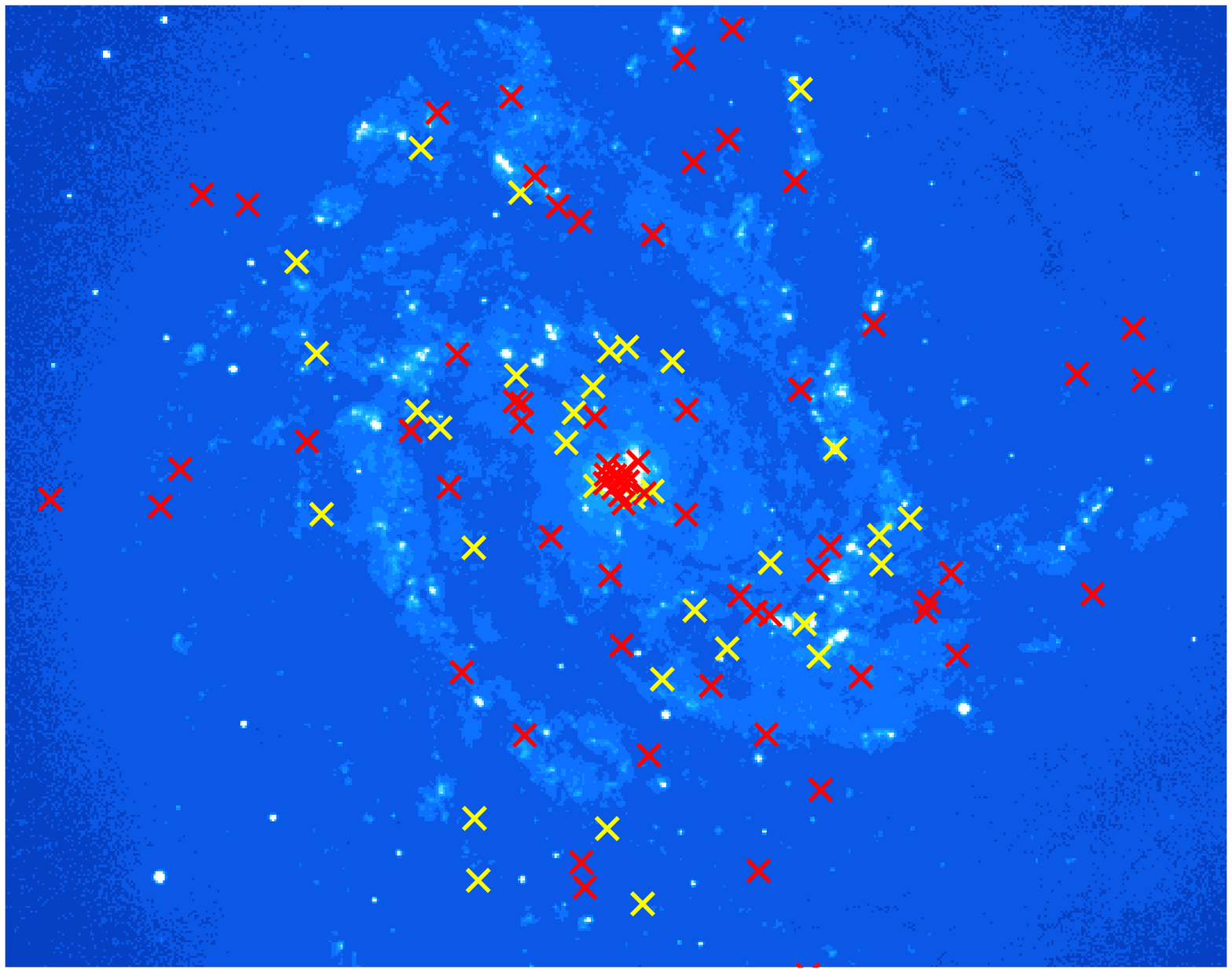}
\caption{M83 X-ray sources plotted on the \chandra\ image (left) and U-band
image (right).  The red X's are sources whose colors make them
good candidates for LMXBs; the yellow X's are soft sources
(possible SNRs)}
\label{fig:opt_and_x-ray}
 \end{figure}

The spatial distribution of the soft (yellow X's) and ``LMXB''
(red X's)  sources is shown in
Figure~\ref{fig:opt_and_x-ray} for M83.  The \chandra\ X-ray image is
shown on the left, and the U-band optical image on the right.  Both
``LMXB'' and soft sources have a tendency to occur in the spiral
arms.  The main difference between the two distributions, however, is
that the nuclear X-ray sources are almost all sources with ``LMXB''
colors.  In M101 the X-ray sources follow a similar pattern (see also
\citet{pence01}) with both
types of sources following the spiral structure.  There is only a
single  diffuse source at the center of M101 (there is
essentially no bulge in this system) so segregation of central sources
is not seen.
 
\section{Summary and Conclusions}
\label{sec:s_and_c}

In this paper, we show that there is a highly significant difference
in the X-ray colors of sources in bulge and disk systems.  Disk
galaxies have an additional population of soft X-ray sources and a
scattering of hard sources not seen in bulge systems.  The hard disk
sources are probably HMXBs.  The soft disk sources are probably
dominated by SNR, but we cannot rule out a contribution from a new
population of soft, faint, absorbed accretion sources associated with
the young stellar population.

The differences in X-ray colors of sources in nearby disk and bulge
galaxies can be used as a starting point for source classification,
with the color-color diagram approximately separating SNR, LMXB and
HMXB sources.  These conclusions are strengthened by the
location of known thermal SNRs and binary pulsars in the X-ray
color-color diagram.  The luminosities of sources identified as SNR
and HMXBs are consistent with what is observed in the Galaxy
\citep{grimm01}.  The variability characteristics of different classes
of sources should provide additional constraints on their nature.
Accreting binaries are likely to vary stochastically while the flux from thermal SNR should remain constant.  Unfortunately, although
there are two \chandra\ observations for both disk galaxies (M101 and
M83) the second observations are not deep enough to effectively
constrain the variability of the soft sources identified as SNR.

It is not possible to identify a source on the basis of X-ray
color alone.  However, separating sources on the basis of X-ray colors will be very valuable
for population studies.  
Although X-ray colors provide an excellent starting point for the
classification of sources in nearby galaxies, there is some overlap in
source type in the X-ray color-color diagram.  For example, supernova
remnants which are very highly absorbed will move vertically upwards
into the ``LMXB'' part of the color-color diagram.  The population of
soft sources may contain some supersoft sources or X-ray binaries with
soft spectra.  Therefore other
information such as source variability, luminosity, and optical
counterparts as well as position within the galaxy  must be used to complete the classification process.
Deeper \chandra\ images of nearby galaxies are required to
investigate the variability and spectral  properties of the different source classes
outlined here.

\section{Acknowledgments}

Thanks to Harvey Tananbaum for improving 
the first draft of this paper, Randall Smith for his help with SLANG
scripting and Rosanne Di Stefano and Roberto Soria for stimulating
discussions.  This work was supported by NASA
contract NAS 8-39073 (CXC) and GO1-2029A.

\clearpage

\begin{deluxetable}{llrrcccclcc}
\label{tab:prop}
\tabletypesize{\scriptsize}
\rotate
\tablecaption{Properties of observations and galaxies \label{tab:prop}}
\tablewidth{0pt}
\tablehead{
  \colhead{Galaxy} & \colhead{Hubble} & \colhead{ObsID} & \colhead{Date} &
    \colhead{Exposure} & \colhead{$n_{\rm H}$} & 
    \colhead{Distance} & \colhead{$L_{min}$} & \colhead{N(bkg)} &
\colhead{N(src)}  \\
  \colhead{} & \colhead{type} & \colhead{} & \colhead{} & 
  \colhead{(sec)} & \colhead{($10^{20}$ $cm^{-2}$)} & 
  \colhead{(Mpc)} & \colhead{} & \colhead{} & \colhead{} }
\startdata
NGC 1291 & S0/a & 2059 & 2000-11-07 & 22906 & 2.24 & 8.6 & 4.4e37 & 3 & 44
\\
M83 & Sc & 793 & 2000-04-29 & 48562 & 3.70 & 4.7 & 7.1e36 & 6 & 122 \\
M83 & Sc & 2064 & 2001-09-04 & 7207 & 3.70 & 4.7 & 4.8e37 & 1 & 38 \\
M31 & SA & 309 & 2000-06-01 & 4942 & 6.68 & 0.7 & 1.7e36 & 1 &  \\
NGC 4697 & E6 & 784 & 2000-01-15 & 39013 & 2.14 & 23.3 & 2.0e38 & 5 & 66
\\
M101 & SAB & 934 & 2000-03-26 & 97602 & 1.15 & 5.4 & 3.7e36 & 10 & 113 \\
M101 & SAB & 2065 & 2000-10-29 & 9583 & 1.15 & 5.4 & 3.8e37 & 1 & 10 \\
\enddata
\end{deluxetable}

\clearpage

%\begin{figure}
%\epsscale{0.75}
%\plotone{M83_stack_ss_and_bg.ps}
%\caption{The composite spectrum of M83 soft sources (top) compared to
%a typical background region (taken from the spiral arms)}
%\label{fig:M83_scom}
% \end{figure}

%\begin{figure}
%\epsscale{0.75}
%\plotone{M83_ss_fit_and_resid.ps}
%\caption{Best fit model to M83 soft source composite spectrum (top)
%and residuals (bottom)}
%\label{fig:M83_scom_fit}
% \end{figure}

\end{document}